\documentclass[aps,pra,twocolumn,showpacs,floatfix]{revtex4}
\usepackage{epsfig}
\usepackage{graphicx}
\usepackage{dcolumn}
\begin{document}
\title{Relativistic many-body calculations of electric-dipole 
matrix elements, lifetimes and polarizabilities in rubidium.}
\author{M. S. Safronova}
\author{Carl J. Williams}
\author{Charles W. Clark}
 \affiliation{
Physics Laboratory,
National Institute of Standards and Technology,
Technology Administration,
U.S. Department of Commerce, 
Gaithersburg, Maryland 20899-8410}
\date{\today} 
\begin{abstract}
Electric-dipole matrix elements for $ns-n^{\prime}p$, $nd-n^{\prime}p$,
and $6d-4f$ transitions in Rb are calculated using a relativistic all-order
method.  A third-order calculation is also carried out for these
matrix elements to evaluate the importance of the high-order many-body
perturbation theory contributions. The all-order matrix elements are used to
evaluate lifetimes of $ns$ and $np$ levels with $n=6, 7, 8$ and 
$nd$ levels with $n=4, 5, 6$  for comparison with experiment and to 
provide benchmark values for these lifetimes. The 
dynamic polarizabilities are calculated for $ns$ states of rubidium. 
The resulting lifetime and polarizability values are compared with 
available theory and experiment.
\end{abstract}
\maketitle
\section{Introduction}
A recent proposal \cite{gate} for quantum computation utilizes the 
ground hyperfine states of a neutral atom as qubits, and realizes 
two-qubit quantum gates by conditional excitations to 
Rydberg states. The atoms are
confined to the sites of an 
optical lattice, and the optical potential seen by the atom depends 
on the dynamic polarizability of the atom. Therefore, the atom 
will generally move in a different lattice potential when it is 
excited to the Rydberg state during the gate operation, which may 
cause motional heating and lead to decoherence. In a recent paper
\cite{us}, we proposed two solutions to this problem by matching   
the ac polarizabilities of the atom in the ground and Rydberg 
states.  In the first scheme, the polarizabilities are matched 
 for the specific values of the lattice photon frequency between the
 resonances. In the second scheme, some accidental matches between 
 transition energies are used to match the ground state polarizability with the 
 polarizability of the selected Rydberg states ($11p$ and $15p$ in the case
of 
 Rb). The elimination of motional decoherence is important 
 in helping to design a high-fidelity two-qubit gate 
 capable of meeting the error threshold for scalable quantum 
 computation.
 
  Despite the existence of high-precision 
 measurements of the primary transition 
 electric-dipole matrix elements in alkali-metal atoms,
  accurate experimental data for other transitions are lacking with the 
  exception of a very few transitions in Cs owing to the study of 
  parity nonconservation. In light of the importance of the
 atomic calculations for the quantum logic gate scheme with conditional 
 excitations to Rydberg levels, we have calculated the electric-dipole
matrix elements 
 for subsequent evaluation of lifetimes and polarizabilities
 for a number of Rb levels. We note that the interest in this 
 particular logic gate scheme with neutral atoms results from 
 its potential for fast (sub-microsecond) gate operations.

We have performed all-order calculations of the $7s-np$, $8s-np$,
 $4d-np$, $5d-np$, and $6d-np$ electric-dipole matrix elements
 with $n=5, 6, 7$ and $8$, and of the $6d-4f$ matrix elements.
These results are combined with previous all-order calculations of the 
$5s-np$ and $6s-np$ matrix elements \cite{relsd} to obtain 
values  of the lifetimes of the $6s$, $7s$, $8s$, $6p$, $7p$, $8p$, 
$4d$, $5d$, and $6d$ levels. The lifetime of the $6d_{3/2}$
level is of special interest owing to the large discrepancies between 
existing experiments \cite{4a,4b}.  The third-order matrix-element
calculation  
has also been carried out to evaluate the importance of the higher-order 
contributions. The dynamic polarizabilities of the Rb
$ns$ states are also calculated for both low-lying and Rydberg levels.
 The evaluation of the accuracy of the polarizability calculations is 
 conducted including a comparison between our data and
  other theory and experiment. 

\section{Electric-dipole matrix elements}
 The electric-dipole matrix elements for the $5s-5p$ 
  transitions were measured to high accuracy in \cite{2} and 
  matrix elements for $ns-n^{\prime}p$  
  transitions  with $n=5,6$ and $n^{\prime}=5,6,7,8$ were calculated 
  using a single-double (SD) all-order method in Refs.~\cite{relsd,th}.
In the present work, we have calculated matrix elements for $7s-np$,
$8s-np$,
 $4d-np$, $5d-np$, and $6d-np$  transitions
 with $n=5, 6, 7$ and $8$ as well as $4f-6d$
  transitions  using 
  SD all-order method \cite{sd}. Very briefly, the wave function is 
  represented as an expansion 
  \begin{eqnarray}
 |\Psi_v \rangle &= &\left[ 1 + \sum_{ma} \, \rho_{ma}
a^\dagger_m a_a + \frac{1}{2} \sum_{mnab} \rho_{mnab} a^\dagger_m
a^\dagger_n a_b a_a +
 \right. \nonumber \\
& + &\left. \sum_{m \neq v} \rho_{mv} a^\dagger_m a_v + \sum_{mna}
\rho_{mnva} a^\dagger_m a^\dagger_n a_a a_v \right]|
\Phi_v\rangle  , \label{eq1}
\end{eqnarray}
where $\Phi_v$ is the lowest-order atomic state function, which is
taken to be the {\em frozen-core} Dirac-Hartree-Fock (DHF)
    wave function
 of a state $v$; $a^\dagger_i$ and $a_i$ are
creation and annihilation operators, respectively. 
  \begin{table*}
\caption{\label{tab1}Absolute values of  all-order SD electric-dipole matrix
elements 
for the $np-n^{\prime}s$ transitions in Rb in  atomic units. The 
lowest-order (DHF) and third-order values all also given.  All-order values
for $5s-np$ and $6s-np$ transitions are 
from Ref.~\protect\cite{relsd}. The corresponding reduced oscillator
strengths are given in rows labeled $f$.
All-order matrix elements and experimental energies   are used
in calculation  of 
oscillator strengths.} 
 \begin{ruledtabular}  
\begin{tabular}{lcccccccc}         
   &  $5s-5p_{1/2}$ & $5s-6p_{1/2}$ & $5s-7p_{1/2}$& $5s-8p_{1/2}$      &
$5s-5p_{3/2}$ &$5s-6p_{3/2}$&  $5s-7p_{3/2}$& $5s-8p_{3/2}$   \\    [0.1pc]
\hline
    DHF         &      4.819  &    0.383      &   0.142     &    0.078  &
6.807    &  0.606       & 0.237        &0.136            \\
    Third order &      4.181  &    0.363      &   0.130     &    0.069  &
5.899    &  0.583       & 0.224        &0.125            \\
    All order   &      4.221   &    0.333      &   0.115     &    0.059 &
5.956    &  0.541       & 0.202        &0.111            \\  
    $f_{5s\rightarrow np}$          &       0.3404  &  0.0040  &    0.0006&
0.0002&    0.6905&    0.0106&    0.0017&    0.0006\\[0.1pc]
       &$6s-5p_{1/2}$&   $6s-6p_{1/2}$& $6s-7p_{1/2}$&   $6s-8p_{1/2}$
&$6s-5p_{3/2}$ & $6s-6p_{3/2}$&  $6s-7p_{3/2}$& $6s-8p_{3/2}$   \\[0.1pc]
    DHF         &      4.256  &   10.286       & 0.976        &0.375    &
6.187   &   14.457     &  1.498        &0.597            \\
    Third order &      4.189  &    9.584       & 1.050        &0.420    &
6.115   &   13.447     &  1.610        &0.668            \\
    All order   &      4.119  &    9.684       & 0.999        &0.393    &
6.013   &   13.592     &  1.540        &0.628            \\
       $f_{6s\rightarrow np}$        &      -0.1946  &  0.5101  &    0.0117&
0.0023&   -0.4019&    1.0267&    0.0279&    0.0058\\[0.1pc]
     &$7s-5p_{1/2}$ &  $7s-6p_{1/2}$&  $7s-7p_{1/2}$ & $7s-8p_{1/2}$
&$7s-5p_{3/2}$& $7s-6p_{3/2}$& $7s-7p_{3/2}$&  $7s-8p_{3/2}$    \\ [0.1pc]
    DHF        &       0.981  &    9.360      & 17.612         & 1.801  &
1.393  &   13.552     &  24.708      & 2.728             \\
    Third order&       0.952  &    9.304      & 16.679         & 1.944  &
1.347  &   13.517     &  23.349      & 2.943             \\
    All order  &       0.954  &    9.189      & 16.844         & 1.865  &
1.352  &   13.353     &  23.587      & 2.833             \\
      $f_{7s\rightarrow np}$        &       -0.0190  & -0.3330  &    0.6565&
0.0186&   -0.0375&   -0.6821&    1.3170&    0.0432 \\[0.1pc]
             &$8s-5p_{1/2}$& $8s-6p_{1/2}$& $8s-7p_{1/2}$&$8s-8p_{1/2}$
&$8s-5p_{3/2}$ & $8s-6p_{3/2}$&$8s-7p_{3/2}$& $8s-8p_{3/2}$     \\[0.1pc]
    DHF         &       0.514  &     1.922     &  16.151       &  26.817&
0.727   &    2.705     & 23.343      &37.577             \\
    Third order &       0.500  &     1.839     &  16.162       &  25.587&
0.705   &    2.578     & 23.428      &35.770             \\
    All order   &       0.504  &     1.853     &  15.982       &  25.831&
0.710   &    2.600     & 23.171      &36.123         \\
    $f_{8s\rightarrow np}$        &         -0.0063  & -0.0278  &   -0.4701&
0.7987&   -0.0124&   -0.0539&   -0.9595&     1.5993\\ 
 \end{tabular}   
\end{ruledtabular}
 \end{table*}              
  \begin{table}[h]
\caption{\label{tab2} Absolute values of all-order SD electric-dipole 
matrix elements 
for the $np-n^{\prime}d$ and $6d-4f$ transitions in Rb in  atomic 
units. The lowest-order and third-order values all also given.
In the cases where lowest order and SD data are of different sign relative 
signs are shown.} 
 \begin{ruledtabular}  
\begin{tabular}{lcccc}  
$n$              &$5$  & $6$ &  $7$ & $8$\\ [0.1pc] \hline

              & \multicolumn{4}{c}{$4d_{3/2}-np_{1/2}$}\\ [0.1pc]
DHF           &    9.046            &     6.725            &   
1.181              &  0.578\\ Third order   &    8.092            
&     5.289            &   1.119              &  0.566\\ All order     
&    7.847            &     4.717            &   1.054              
&  0.541  \\[0.1pc] 
              &  \multicolumn{4}{c}{$4d_{3/2}-np_{3/2}$}\\[0.1pc]
DHF      &    4.082            &     2.955            &   0.534

&  0.262\\ Third order   &    3.655            &     2.307            
&   0.502              &  0.255\\All order     &    3.540            
&     2.055            &   0.470              &  0.242\\[0.1pc] 
              &  \multicolumn{4}{c}{$4d_{5/2}-np_{3/2}$}\\[0.1pc]   
DHF           &   12.241            &     8.829            &   
1.601              &  0.787\\ Third order   &   10.964            
&     6.915            &   1.505              &  0.764\\ All order     
&   10.634            &     6.184            &   1.411              
&  0.726\\ \hline  
              &  \multicolumn{4}{c}{$5d_{3/2}-np_{1/2}$}\\[0.1pc]  
DHF          &     0.244          &     18.701          &      
13.639         &        2.660\\ Third order  &     1.220          
&     18.241          &      10.600         &        2.573\\ All 
order    &     1.616          &     18.106          &       9.768         
&        2.400\\[0.1pc] 
               &  \multicolumn{4}{c}{$5d_{3/2}-np_{3/2}$}\\[0.1pc]   
DHF          &     0.157          &      8.443          &        
5.983        &        1.198\\ Third order  &     0.607          &      
8.232          &        4.609        &        1.146\\ All order    
&     0.787          &      8.160          &        4.242        &        
1.064\\[0.1pc] 
              &  \multicolumn{4}{c}{$5d_{5/2}-np_{3/2}$}\\[0.1pc]   
DHF          &     0.493          &     25.340          &       
17.884        &         3.592\\ Third order  &     1.821          
&     24.695          &       13.843        &         3.435\\ All 
order    &     2.334          &     24.491          &       12.798        
&         3.201 \\[0.1pc] \hline      
                 &  \multicolumn{4}{c}{$6d_{3/2}-np_{1/2}$}\\[0.1pc]  
DHF         &      0.512          &      -0.254          &       
31.349       &           22.584\\ Third order &      1.076          
&      1.375          &       31.563       &           17.647\\ 
All order   &      1.180          &      1.989          &       
31.422       &           16.631\\[0.1pc] 
                  &  \multicolumn{4}{c}{$6d_{3/2}-np_{3/2}$}\\[0.1pc]      
DHF         &      0.255          &       -0.028         &       
14.158       &             9.901\\ Third order &      0.513          
&       0.732         &       14.242       &             7.662\\ 
All order   &      0.558          &       1.012         &       
14.161       &             7.215\\  [0.1pc] 
                   &  \multicolumn{4}{c}{$6d_{3/2}-np_{3/2}$}\\[0.1pc]  
DHF          & 0.778  & -0.047 &   42.500   &   29.603\\
Third order &  1.532  & 2.184 &   42.709   &   23.041\\ 
   All order&  1.658  & 2.974 &   42.481   &   21.784\\
   \hline
     & $6d_{3/2}-4f_{5/2}$  &$6d_{5/2}-4f_{5/2}$ &$6d_{5/2}-4f_{7/2}$&
\\[0.1pc] 
     \hline
DHF         &  6.109  & 1.642  &  7.343&\\
Third order &  9.150  & 2.443  & 10.924&\\
   All order&  9.938  & 2.642  &  11.813&\\      
 \end{tabular}   
\end{ruledtabular}
 \end{table}
     The index $a$ is used
  to represent core states and indices $m$ and $n$ indicate excited states.

The quantities
 $\rho_{ma}$ and $\rho_{mv}$ are single core and valence excitation
 coefficients and  the quantities $\rho_{mnab}$ and $\rho_{mnva}$ are double
core
  and valence excitation coefficients, respectively.
   We obtain the equations for the excitation coefficients by substituting
the 
   above wave function into the many-body Schr\"{o}dinger equation. The
equations are 
   solved iteratively and the resulting expansion coefficients are used to
calculate matrix
   elements. In such a procedure, certain classes of MBPT terms are summed to
all orders. 
   However, the restriction of the expansion (\ref{eq1}) to the single and
double excitations
   leads to some missing terms in the expression for the matrix elements
starting from the
   fourth order. As it is important to understand how significant
higher-order
   corrections are for the transitions under consideration, we also conduct
separate 
   third-order calculation of these matrix elements (including random-phase
approximation
  (RPA) contributions iterated to all orders). The third-order calculation
    follows that of Ref.~\cite{ADNDT}. Such a calculation is of
    lower accuracy than the 
   all-order one but the difference in third-order and all-order values
provides an
  estimate of the importance of the higher-order contributions. Also, the
breakdown of 
  both all-order and third-order calculations to different contributions
yields
  information regarding relative importance of the specific terms and
possible 
  large cancellations between different terms. Such information can be used
to approximate 
  some of the omitted  contributions and further estimate the uncertainty of
the 
  all-order calculation.  We should note that despite the fact that the
entire
  third-order contribution is contained in the all-order result, the
extraction of the 
  third-order part from the all-order calculation is not a straightforward
  task (see Ref.~\cite{th}
  for the correspondence of the terms), so a separate calculation is made to
  obtain third-order values. The separate calculation also allows us to
iterate
   RPA contribution to
  all orders. 
    \begin{table*}
\caption{\label{tab3} Contributions to third-order  and all-order
electric-dipole reduced matrix elements (a.u.) for selected 
transitions in Rb.} 
 \begin{ruledtabular}  
\begin{tabular}{lrrrrrrrr}
\multicolumn {1}{c}{}&
\multicolumn {1}{c}{$5s-5p_{1/2}$}&
\multicolumn {1}{c}{$5s-7p_{1/2}$}&
\multicolumn {1}{c}{$7s-5p_{1/2}$}&
\multicolumn {1}{c}{$7s-7p_{1/2}$}&            
\multicolumn {1}{c}{$5d_{3/2}-5p_{1/2}$}&  
\multicolumn {1}{c}{$5d_{3/2}-7p_{1/2}$}&
\multicolumn {1}{c}{$6d_{3/2}-6p_{1/2}$}&
\multicolumn {1}{c}{$6d_{3/2}-4f_{5/2}$}\\
        \hline
 \multicolumn {9}{c}{Third order}\\ [0.2pc]
  DHF   &   4.819&   0.149&  0.981 & 17.612&  0.244& 13.639 & -0.254&
6.109\\
  RPA   &  -0.213&  -0.060&  0.015 & -0.024&  0.101&  0.030 &  0.045&
0.002\\
  BO    &  -0.419&   0.040& -0.038 & -0.899&  0.888& -3.048 &  1.589&
3.041\\
  SR    &   0.027&   0.009& -0.004 &  0.002& -0.013& -0.004 & -0.005&
0.001\\
  Norm  &  -0.033&  -0.001& -0.002 & -0.012& -0.001& -0.016 &
0.000&-0.004\\
  Total &   4.181&   0.130&  0.952 & 16.679&  1.220& 10.600 &  1.375&
9.150\\[0.2pc]
 \hline                                             
 \multicolumn {9}{c}{All order}\\[0.2pc]
  DHF    &  4.819&  0.142 &  0.981  &   17.612 &    0.244 &   13.639 &
-0.254  &   6.109\\
  Term $a$ & -0.234& -0.064 &  0.014  & -0.025 &    0.102 &    0.036 &
0.047  &   0.005\\
  Term $c$ & -0.375&  0.009 & -0.077  & -0.648 &    1.575 &   -3.372 &
2.752  &   4.393\\
  Term $d$ &  0.039&  0.017 &  0.056  &  0.149 &   -0.197 &   -0.024 &
-0.439  &  -0.089\\
  Other  &  0.039&  0.012 & -0.006  &    0.005 &   -0.025 &   -0.004 &
-0.009  &   0.002\\
  Norm   & -0.068& -0.002 & -0.014  &   -0.248 &   -0.083 &   -0.506 &
-0.109  &  -0.483\\
  Total  &  4.220&  0.115 &  0.954  &   16.844 &    1.616 &    9.768 &
1.989   &  9.938\\[0.2pc]
 \end{tabular}  
\end{ruledtabular}
 \end{table*}
 
  We use the 
system of atomic units, a.u., in which $e, m_{\rm e}$, $4\pi 
\epsilon_0$ and the reduced Planck constant $\hbar$ have the 
numerical value 1.  Polarizability in a.u. has the dimensions of 
volume, and its numerical values presented here are thus measured 
in units of $a^3_0$, where $a_0\approx0.052918$~nm is Bohr radius.
The atomic units for $\alpha$ can be be converted to SI units via
 $\alpha/h$~[Hz/(V/m)$^2$]=2.48832$\times10^{-8}\alpha$~[a.u.], where
 the conversion coefficient is $4\pi \epsilon_0 a^3_0/h$ and 
 Planck constant $h$ is factored out. The atomic unit of frequency $\omega$
 is $E_h/\hbar\approx4.1341\times10^{16}$~Hz, where $E_h$ is Hartree
 energy. 
 
   The third-order and all-order  results for $ns-n^{\prime}p$ transitions
  with $n,n^{\prime}=5, 6, 7$ and $8$ are summarized in Table~\ref{tab1}
 together with the lowest-order Dirac-Hartree-Fock (DHF) values.
 The corresponding reduced oscillator strengths calculated using 
 formula \cite{1}
 \begin{equation}
 f_{ab}=-\frac{30.3756}{(2j_a+1)\lambda} |\langle a\|D\|b\rangle|^2
 \label{f}
 \end{equation}
 are listed in the rows labeled $f_{ns\rightarrow n^{\prime}p}$.
 In Eq.~(\ref{f}), 
 $D$ is the dipole operator and $\lambda$ is a transition wavelength in 
 nm.   The all-order 
 matrix elements and experimental energies from \cite{exp} are used in 
 the oscillator strength calculation. The sum of the $f_{5s\rightarrow np}$ 
 oscillator strengths with $n=5,6,7,8$ slightly exceeds one because the
contributions 
 to the sum rule $\sum_{n}f_{5s\rightarrow np}=1$ from the transitions with 
 $n=2, 3, 4$ are negative.
 
The results for the $nd-n^{\prime}p$ matrix elements with $n=4, 5, 6$
and the $n^{\prime}=5,6,7,8$ and $6d-4f$ matrix elements are
summarized in Table~\ref{tab2}. 
 The transitions 
containing the fine structure components, such as $ns-5p_{1/2}$ and
$ns-5p_{3/2}$
pairs, for example, have very similar relative correlation contributions so
we will
omit the angular momentum $j$ subscript in the subsequent discussion.
 We find that the relative correlation correction
contribution varies very significantly with the transition. It is small,
below 7\%,
 for all $6s-np$, $7s-np$ and $8s-np$ transitions and for some
$nd-n^{\prime}p$
 transitions. 
In those situations where the third order and  all-order values 
differ by less than a few percent we expect the dipole matrix 
elements to be accurate to at least 2\%. 
 
We should note that in some cases it is possible  that such good agreement
of the 
third-order and all-order values is fortuitous and may result 
from accurate cancellations of the 
high-order terms. We address this issue below in more detail. 
 
  In Ref.~\cite{relsd}, the all-order values for the primary 
  transitions were found to agree with a recent high-accuracy measurement
  \cite{2} to within 0.2\%--0.35\%. The accuracy of the $5s-6p$, $5s-7p$,
and $5s-8p$
  matrix elements is, however, substantially lower since they are
relatively small 
  matrix elements with significant relative correlation contributions. 
  To investigate the correlation 
  contributions in more detail, we give a breakdown of both third-order 
  and all-order calculations for eight selected transitions
  in Table~\ref{tab3}. The third-order contributions are 
  separated to random-phase-approximation (RPA), Brueckner-orbital (BO),
  structural radiation (SR) and normalization (Norm)
  corrections  following Ref.~\cite{ADNDT}. For the all-order contributions,
three
  terms ($a$, $c$ and $d$) are listed separately and the other 17 terms are
grouped
   together in row labeled ``Other''. The derivation and  expressions for these
terms are
   given in Refs.~\cite{sd,th}.  A normalization correction is
    given in row labeled ``Norm''.
   As it was noted earlier, there is no straightforward correspondence of
the 
   all-order and third-order breakdown (full description is given in
\cite{th}), but 
   term $a$ partly corresponds to RPA-like corrections and term $c$ to
BO-like
   corrections. Term $d$ is normally small with exception of some
transitions.
    It is 
   quadratic in single-valence excitation coefficients $\rho_{mv}$ and,
therefore, 
      contains only fifth and higher-order terms. 
   We find strong cancellations between BO and RPA terms for 
   $5s-6p$, $5s-7p$, and $5s-8p$ matrix elements. The difficulty of the 
   calculation of such matrix elements has been described before for the
case of Cs
   \cite{relsd,cs,Al}. However, we also find that term $d$, which is missing
    entirely from third-order calculation is relatively large for $7s-5p$
and $7s-7p$
    matrix elements. It is rather puzzling since we observe 
    very good agreement of third-order
    and all-order results for these transitions. The possible explanation is
    accurate cancellation of the high-order terms for these transitions. 
    We note that term $d$ is not significant for the $6s-5p$ transition, for
which some 
    conclusion of accuracy may be drawn from the comparison with similar 
    Cs transition matrix elements (which are measured to quite good
accuracy).
    Accurate measurements of these matrix elements would be very useful in
developing further 
    understanding of this issue. We note that even though this term is
significant, it is 
    not very large, on the order of 5\%. 
    
    In the case of $nd-n^{\prime}p$ transitions, the term $c$ (or BO term
for third-order 
    calculation) dominates. The relative contribution of the correlation
     correction varies 
    with the transition. For $5d-5p$ and  $6d-6p$ transitions,
     the DHF approximation gives a very poor result; the sign of the matrix
element 
     changes when correlation is added for $6d-6p$ matrix elements 
     and the DHF values for $5d-5p$
     matrix elements are about 7 times too low.    
     As expected, we find very large, $25\%-30\%$, differences between 
     third-order and all-order
     calculations for these transitions as the all-order calculation
includes 
     correlation more 
     completely. The  term $c$ can be 
     corrected by including triple excitations or use of the semi-empirical
      scaling described in Refs.~\cite{cs,relsd,th}. 
     To check the validity of the scaling approach, we applied such a method to
the 
     $5d-5p$ matrix elements and compared the resulting ratio with the
experimental 
     measurement \cite{andrei}.
     
   \begin{table}[h]
\caption{\label{tabR} The ratio $R$
of the $5d_{3/2}-5p_{3/2}$ and $5d_{3/2}-5p_{1/2}$
 electric-dipole matrix elements divided by the corresponding values of the 
 $C_1$ reduced matrix elements, where $C_1$ is the normalized spherical
harmonic.
  Comparison
  of the present results calculated in  different approximations with 
  theoretical and experimental values from Ref.~\protect\cite{andrei}. The
corresponding $5d-5p$ matrix
   elements, divided by the corresponding values of the 
 $C_1$ reduced matrix elements are also given.} 
 \begin{ruledtabular}  
\begin{tabular}{lcccc}   \multicolumn{1}{c}{}&
  \multicolumn{3}{c}{$5p_J-5d_{J^{\prime}}$} & $R$ \\
  $J-J^{\prime}$& $3/2-5/2$  &$3/2-3/2$ & $1/2-3/2$ &    \\ 
     \hline
DHF               & 0.3182   & 0.3034 & 0.2115  &  1.434\\ Third 
order       & 1.1756   & 1.1758 & 1.0562  &  1.113  \\ All order         
& 1.5063   & 1.5248 & 1.3995  &  1.089    \\ Scaled            & 
1.2801   & 1.2882 & 1.1707  &  1.100   \\ MBPT
~\protect\cite{andrei}& 1.0238   & 1.0216 & 0.9000  &  1.135    
\\ Expt. \protect\cite{andrei} &         &        &          & 
1.068(8) \\ 
 \end{tabular}   
\end{ruledtabular}
 \end{table}
 In Table~\ref{tabR} we give our values of the ratio $R$
 $$
 R=\frac{\langle 5p_{3/2}\|D\|5d_{3/2}\rangle}{\langle
5p_{1/2}\|D\|5d_{3/2}\rangle}
   \frac{\langle 5p_{1/2}\|C_1\|5d_{3/2}\rangle}{\langle
5p_{3/2}\|C_1\|5d_{3/2}\rangle}
 $$
of the $5p_{3/2}-5d_{3/2}$ and $5p_{1/2}-5d_{3/2}$
 electric-dipole matrix elements divided by the corresponding values of the 
 $C_1$ reduced matrix elements, where $C_1$ is the normalized spherical
harmonic \cite{1}.
 The present results are compared with theoretical and experimental values
from Ref.~\cite{andrei}.
  The corresponding $5d-5p$ matrix
   elements, divided by the corresponding values of the 
 $C_1$ reduced matrix elements are also given.
 The theoretical value of Ref.~\cite{andrei} was obtained using relativistic
  third-order many-body theory \cite{ADNDT}; some higher-order
Brueckner-orbital terms were
 also included, which accounts for the difference with our third-order
result. 
 As noted above, the largest contribution to the $5p-5d$ matrix elements
  calculated using the all-order 
 method comes from the term $c$
 containing single-valence excitation coefficients $\rho_{mv}$. We scale
these
 excitation coefficients $\rho_{mv}$ with the ratio of the experimental
  and corresponding theoretical correlation energies as described in 
 Refs.~\cite{cs,relsd}. We list the results  obtained using such scaling 
 in a row labeled ``Scaled''. We find very significant differences 
 between all high-precision calculations, which indicates that the accuracy of
the all-order
 calculation is around 10-20\% for these matrix elements.
   However, our all-order result for the ratio
$R$  is in much better agreement with experiment \cite{andrei} than
the third-order value
 and theoretical
 calculation  of Ref.~\cite{andrei}. 


\section{Lifetimes}

 We use the resulting all-order matrix elements to 
 calculate the lifetimes of the  $6s$, $7s$, $8s$,
  $6p$, $7p$, $8p$, $4d$, $5d$, and $6d$  levels in Rb
   for the comparison with experiment and
  providing benchmark values for these
 lifetimes. 
 
 The Einstein A-coefficients $A_{vw}$ 
 \cite{1} are calculated using 
 the formula
 \begin{equation}
 A_{vw}=\frac{2.02613\times 10^{15}}{\lambda^3}
 \frac{|\langle v\|D\|w\rangle|^2}
 {2j_v+1}\,s^{-1},
 \end{equation}
 where $\langle v\|D\|w\rangle$ is the reduced electric-dipole 
 matrix element for the transition between states $v$ and $w$ and   
  $\lambda$ is corresponding wavelength in nm. The lifetime
  of the state $v$ is calculated as
   \begin{equation}
\tau_v=\frac{1}{\sum_{w}A_{vw}} \label{eq3} 
 \end{equation}
 and the denominator of Eq.~(\ref{eq3}) gives the radiative width
 of the level $v$.
 
  \begin{table} [ht]
\caption{\label{tab4} Transition energies ($E/hc$) in cm$^{-1}$, 
matrix elements (in a.u.) and corresponding contributions to 
$7p_{3/2}$ radiative width (in MHz).} 
 \begin{ruledtabular}  
\begin{tabular}{lrrrr}
    Transition ($vw$)     &$\langle v\|D\|w\rangle$& $\delta E_{vw}$&
    $A_{vw}$ &$\sum_w A_{vw}$\\
    \hline
 $7p_{3/2}-7s   $    & 23.587&     1559&   1.068&   1.068  \\
 $7p_{3/2}-6s   $    &  1.540&     7736&   0.556&   1.624  \\
 $7p_{3/2}-5s   $    &  0.202&    27870&   0.447&   2.071  \\[0.5pt]
 $7p_{3/2}-5d_{3/2}$ &  4.242&     2169&   0.093&   2.164  \\
 $7p_{3/2}-4d_{3/2}$ &  0.470&     8515&   0.069&   2.234  \\
 $7p_{3/2}-5d_{5/2}$ & 12.798&     2169&   0.847&   3.081  \\[0.5pt]
 $7p_{3/2}-4d_{5/2}$ &  1.411&     8515&   0.623&   3.703  \\
 \end{tabular}   
\end{ruledtabular}
 \end{table}

Several electric-dipole  (E1) transitions contribute to the lifetime 
of each of the  levels considered here. The simplest case is that 
of the $6s$ level, where only $6s-5p_{1/2}$ and $6s-5p_{3/2}$ 
transitions need to be included. To calculate  the lifetime of the 
$8p_{3/2}$ 
 state, we need to include ten E1  transitions and in the case of the $6d_{3/2}$
 level seven E1 transitions are allowed, including the $6d_{3/2}-4f_{5/2}$
 transition. The experimental energies from \cite{exp} are used in the 
 lifetime calculation. We illustrate the importance of the different
channels 
 contributing to 
 the lifetime of the $7p_{3/2}$ state in Table~\ref{tab4}, where we give the
  coefficients $A_{vw}$ for each transition together with the transition
energies 
  from \cite{exp} and corresponding matrix elements from 
  Tables~\ref{tab1}-\ref{tab2}.
 The accumulated sum $\sum_w A_{vw}$ is listed in the last column. 
We find the contribution from the $7p_{3/2}-5d_{5/2}$ transition to be nearly as
large 
as the contribution from the  $7p_{3/2}-7s$ transition. All seven
contributions
need to be included in an accurate calculation. The smallest
contribution to $\sum_w A_{vw}$ comes from $7p_{3/2}-4d_{3/2}$ 
transition and is
around 2\%.   

   \begin{table} [ht]
\caption{\label{tab5} Transition energies $E/hc$ in cm$^{-1}$, 
matrix elements (in a.u.) and  the corresponding contributions to 
$6d_{3/2}$ and $6d_{5/2}$ radiative widths (in MHz).} 
 \begin{ruledtabular}  
\begin{tabular}{lrrrr}
    Transition ($vw$)     &$\langle v\|D\|w\rangle$& $\delta E_{vw}$&
    $A_{vw}$ &$\sum_w A_{vw}$\\
    \hline
 $6d_{3/2}-5p_{1/2}$& 1.180& 16108& 2.948 & 2.948\\
 $6d_{3/2}-6p_{1/2}$& 1.989&  4972& 0.246 & 3.194\\
 $6d_{3/2}-7p_{1/2}$&31.422&   852& 0.309 & 3.504\\[0.5pt]
 $6d_{3/2}-5p_{3/2}$& 0.558& 15870& 0.630 & 4.134\\
 $6d_{3/2}-6p_{3/2}$& 1.012&  4894& 0.061 & 4.195\\
 $6d_{3/2}-7p_{3/2}$&14.161&   817& 0.055 & 4.250\\[0.5pt]
 $6d_{3/2}-4f_{5/2}$& 9.938&  1895& 0.341 & 4.591\\[0.2pc]
 \hline
$6d_{5/2}-5p_{3/2}$&  1.658&   15872&  3.712&  3.712\\
$6d_{5/2}-6p_{3/2}$&  2.974&    4896&  0.351&  4.063\\
$6d_{5/2}-7p_{3/2}$& 42.481&     819&  0.335&  4.398\\[0.5pt]
$6d_{5/2}-4f_{5/2}$&  2.642&    1897&  0.016&  4.414\\
$6d_{5/2}-4f_{7/2}$& 11.813&    1897&  0.322&  4.736\\
 \end{tabular}   
\end{ruledtabular}
 \end{table}
 We list the contributions to $6d_{3/2}$  and $6d_{5/2}$  lifetimes 
 as well as the corresponding matrix elements and transition energies in 
 Table~\ref{tab5}. 
 The dominant contributions to $6d_{3/2}$  and $6d_{5/2}$
 lifetimes come from the $6d_{3/2}-5p_{1/2}$ and $6d_{5/2}-5p_{3/2}$
 transitions, respectively. The next largest contribution is 14\% for
the  $6d_{3/2}$ lifetime and 7\% for the $6d_{5/2}$ lifetime.
 We find the contributions from $6d_{3/2}-4f_{5/2}$ and $6d_{5/2}-4f_{7/2}$
 transitions to be around 7\% for the corresponding level. 
    
   \begin{table}
\caption{\label{tab6}
Radiative widths (in MHz) and lifetimes (in ns) for $ns$, $np$, and
$nd$ states in Rb. Experimental values are taken from the
compilation in Ref.~\protect\cite{5} unless noted otherwise.
The lifetimes obtained using \textit{ab initio} all-order matrix elements
and scaled all-order values are listed in columns $\tau_v$ and
$\tau_v^{\text{sc}}$,
respectively.}
 \begin{ruledtabular}
\begin{tabular}{lrrrrr}
  State     &$\sum_w
A_{vw}$&$\tau_v$&$\tau_v^{\text{sc}}$&$\tau_v^{\text{expt}}$&\\
    \hline
     $6s$   & 21.761 & 46.0&  45.4&  46(5)&\\
     $7s$   & 11.318 & 88.4&  88.3&  88(6)&\\
            &        &     &      &  91(11)&\\ [1pt]
     $8s$   &  6.201 &161.3& 161.8& 161(3) &\protect\cite{4a} \\
            &        &     &      &  154(7)&\\
            &        &     &      &   153(8)&\\
 $6p_{1/2}$ &  7.729 & 129 & 123  &  125(4)&\\
            &        &     &      &  131(5)&\\ [0.2pt]
 $7p_{1/2}$ &  3.394 & 295 & 280  & 272(15)&\\
 $8p_{1/2}$ &  1.899 & 527 & 508  &      &\\ [0.2pt]
 $6p_{3/2}$ &  8.461 & 118 & 113  & 112(3)&\\
 $7p_{3/2}$ &  3.703 & 270 & 258  & 246(10)&\\
            &        &     &      & 233(10)&\\[0.2pt]
 $8p_{3/2}$ &  2.056 & 486 & 471  &  400(80)&\\ [0.5pt]
 $4d_{3/2}$ & 11.478 &  87 &  83.5& 86(6)&\\
 $5d_{3/2}$ &  5.193 & 193 & 243  & 205(40)&\\
 $6d_{3/2}$ &  4.591 & 218 & 263  &256(4)&\protect\cite{4a}\\
            &        &     &      &298(8)&\protect\cite{4b}\\ [0.5pt]
$4d_{5/2}$  & 10.674 &  94 &  90  & 94(6)&\\
$5d_{5/2}$  &  5.340 & 187 & 235  & 230(23)&\\
$6d_{5/2}$  &  4.736 & 211 & 252  &249(5)&\protect\cite{4a}\\
 \end{tabular}
\end{ruledtabular}
 \end{table}

The results for the $ns$, $np$ and $nd$ lifetimes obtained using
all-order matrix elements are compared with experimental values 
from Refs.~\cite{5,4a,4b} in Table~\ref{tab6}. We omit the effect 
of the blackbody radiation in our calculations. In Ref.~\cite{5}, 
it was estimated  to be small for the levels  considered here with 
the exception of the $8p$ levels. It is still far below 
experimental uncertainty of the $8p$ level lifetime. The results 
agree with experiment within the experimental precision for  $6s$, 
$7s$, $8s$, $6p$, and $4d$ levels. The values of the $7p$, $8p$, 
$5d$, and $6d$ lifetimes are in good agreement with experiment. 
The lower accuracy of the calculations is expected for these 
levels owing to the lower accuracy of the $np-n^{\prime}d$ matrix 
elements with comparison to most of the $np-n^{\prime}s$ ones due 
to large correlation corrections for these matrix elements. It 
should be noted that the experimental uncertainties of $7p$, $8p$, 
and $5d$ 
 lifetimes are rather large (up to 20\%) and two recent experiments for the 
 $6d_{3/2}$ level give substantially different values: 256(4)~ns and
298(8)~ns.   
 Our result for $6d_{3/2}$ lifetime  (263~ns) is
  very close to  the first measurement, 
 however, large uncertainties in the correlation contributions 
 to the relevant transitions do not allow to rule out 10\% larger value
 of this lifetime. 
      
 The accuracy of the theoretical calculations for these levels
  may be improved by including some triple-excitation contributions 
  into the corresponding wave functions. 
  
  We estimate some 
  omitted higher-order contributions using the scaling of the single
excitation
  coefficients $\rho_{mv}$ described above. The contributions
   containing these single-excitation coefficients are dominant for  
   all of the transitions needed for the calculation of lifetimes in 
   Table~\ref{tab6}. The lifetimes obtained using  scaled all-order
    matrix elements are listed in column $\tau_v^{\text{sc}}$.
    We find that the scaled data agree with experimental values 
  within the experimental uncertainty for all levels with exception
  of $7p_{3/2}$ and $6d_{3/2}$, where the theoretical values are
   just outside the experimental uncertainties.    

\section{Polarizabilities}
 The valence part of the ac polarizability of an alkali-metal atom in $ns$
state
 can be calculated using the formula 
 \begin{eqnarray}
\alpha_v^{ns}(\omega)&=&\frac{1}{3}\sum_{n^{\prime}}
\left(
\frac{(E_{n^{\prime}p_{1/2}}-E_{ns})
\langle n^{\prime}p_{1/2}\|D\| ns \rangle^2 }
{(E_{n^{\prime}p_{1/2}}-E_{ns})^2-\omega^2}\right.  \nonumber \\
&+&\left. \frac{(E_{n^{\prime}p_{3/2}}-E_{ns})
\langle n^{\prime}p_{3/2}\|D\| ns \rangle^2 }
{(E_{n^{\prime}p_{3/2}}-E_{ns})^2-\omega^2}
\right),
\label{eq2}
\end{eqnarray}
where $D$ is the dipole operator.  
 In this formula,  $\omega$ is 
assumed to be at least several linewidths off resonance with the 
corresponding transition. 

The core contribution to the polarizability, calculated in the 
DHF approximation  is found to be small for Rb
(9.3~$a_0^3$) and is weakly dependent on $\omega$ in the frequency 
range considered here.  The static value for the polarizability of 
Rb$^+$ calculated in the random-phase approximation (RPA) 
\cite{der,jkh} is 9.1 $a_0^3$, close to the value of 9.0 $a_0^3$ 
obtained by Johansson \cite{joh} from analysis of the 
observed term values of nonpenetrating Rydberg states. The 
accuracy of the RPA approximation for the core polarizability  is 
estimated to be 5~\% in Ref.~\cite{relsd}. We use the RPA value 
for the core polarizability of Rb$^+$ as a baseline, and adjust it 
to account for valence electron (using Eq.(\ref{eq2}) with 
$n^{\prime}=2,3,4$) and the frequency dependence by using DHF calculations. 
The RPA and DHF values differ by only 2~\%. The correction to the 
core polarizability owing to the presence of the valence electron 
is very small, it is only $-0.3a^3_0$ for $5s$  state in DHF 
approximation.  

\begin{figure}
\includegraphics[scale=0.45]{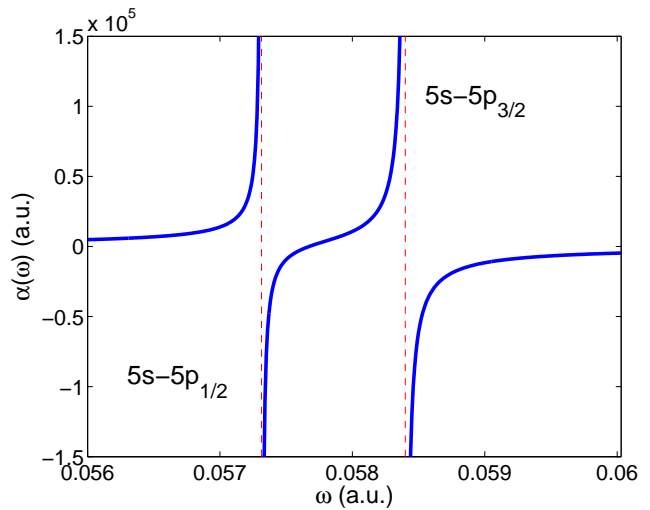}   
\caption{ \label{fig1} Dynamic polarizability $\alpha(\omega)$ for the
ground state 
   of Rb in a.u.   }             
\end{figure}

First, we describe the calculation of the ground state 
polarizability, $\alpha_{5s}(\omega)$. The expression (\ref{eq2}) 
converges rapidly with $n^{\prime}$ so the contribution with $n^{\prime}=5$ is 
dominant. We use experimental $5s-5p$ matrix elements from 
\cite{2}, 
  all-order matrix elements from Ref.~\cite{relsd}, and 
  experimental energies from \cite{exp} to evaluate the  expression of
Eq.~(\ref{eq2})
  with  $n^{\prime}=5,6,7,8$. The contribution to the ground state polarizability
from states 
  with $n^{\prime}>8$ is very small, $0.2a^3_0$ in DHF approximation. 
  We plot the ground state  frequency-dependent polarizability 
  in the vicinity of the $5s-5p_{1/2}$ and $5s-5p_{3/2}$ resonances in
Fig.~\ref{fig1}.
  The behavior within the few linewidths from the resonances is not shown
and the 
  exact placement of the resonances is indicated  by the vertical dashed
lines.                                                              
   As we see from Fig.~\ref{fig1}, $\alpha_{5s}(\omega)$ 
   changes sign between two resonances. We determined that
$\alpha_{5s}(\omega)=0$
   at  $\lambda_{vac}=790.032(8)$~nm \cite{us}, where the uncertainty
results
   from the uncertainty of the polarizability calculation. This crossing
point is of interest 
   for the optical lattice experiments with Rb as the atoms will no longer
be 
   trapped at this wavelength.

We give the breakdown of the different contributions to the ground 
state polarizability for one particular frequency, 
$\omega=0.04298$~a.u., corresponding to $\lambda=1.06\mu$m 
\cite{wavelength_comment}, in Table~\ref{tab7}. The comparison 
with other theory \cite{msd} and experiment \cite{Bonin} is also 
given. Our value is in good agreement with the result from 
\cite{msd}. The discrepancy is due to our use of more accurate 
values for the electric-dipole matrix elements. 
 Our result is just outside the range of uncertainty
of the value $\alpha=769 \pm 61~a_0^3$ inferred by
Bonin and Kadar-Kallen from an atomic deflection 
experiment \cite{Bonin}. 
  \begin{table}                                     
\caption{Contributions to ground state dynamic polarizability of Rb in
$a^3_0$
for $\omega=0.04298$~a.u.
\label{tab7}} 
\begin{ruledtabular}  
\begin{tabular}{lrr}  
    \multicolumn{1}{c}{}& 
   \multicolumn{1}{c}{Contribution}&    
    \multicolumn{1}{c}{Value} \\ 
   \hline
                      & $n^{\prime}=5$ &    682.84 \\
                      & $n^{\prime}=6$ &    1.48  \\
                      & $n^{\prime}=7$ &    0.16 \\
                      & $n^{\prime}=8$ &    0.04 \\   [0.3pc]
                       \hline 
    &$\alpha_v (n^{\prime}=5..8)$ &   684.52(72)  \\
    &$\alpha_v (n^{\prime}>8)$    &   0.17(8)    \\ 
    & $\alpha_{vc}(n^{\prime}=2..4)$&          -0.26(13) \\ 
           & $\alpha_{c}$   &  9.10(45) \\ \hline
Final $\alpha$              & &   693.5(9)\\  
Ref.~\cite{msd} && 711.4\\
Expt. \cite{Bonin}&& 769(61)\\               
\end{tabular}       
\end{ruledtabular}
\end{table}

Next, we describe the calculation of the polarizabilities of the 
Rydberg states. We calculate the polarizabilities of the $ns$  
states of Rb  with $n=8\dots 19$ using  the  DHF approximation; i.e. 
using DHF values for both energies and matrix elements in 
Eq.~(\ref{eq2}).  The summation over $n^{\prime}$  in 
  Eq.~(\ref{eq1})  is
   truncated at $n^{\prime}=23$. The DHF calculations are done on the 
non-linear grid of the 
 form 
 \begin{equation}
 r(i)=r_0(e^{(i-1)h}-1).
 \end{equation}
 To calculate matrix elements and energies of the first few excited states 
 500 grid points are sufficient. This type of grid
 provides a very dense grid near the origin. Thus we needed to confirm
 that there were sufficient points at large $R$ to support the
 Rydberg states. 
 The parameters $r_0$ and $h$ 
 were  chosen to ensure sufficient number of points 
 to calculate $\langle ns\|D\| n^{\prime}p \rangle$ matrix elements
 for high values of $n$ and  $n^{\prime}$.
 However, we found that the accurate DHF values of the matrix elements
  even between  high Rydberg states can be obtained with relatively small
number
  of grid points. For example, the values of the $20s-20p_{3/2}$ matrix element
  obtained with 500 and 40000 point grids (corresponding to approximated 10
and 200 points 
  in the relevant grid section) differ by only 0.05\%.
     The results of the DHF calculation for $\omega$=0.0576645~a.u.
corresponding to 
      $\lambda = 790$~nm are  given in Table~\ref{tab8}. 
     We find that the polarizability values do not change 
     substantially with $n$ for $n>8$. We conduct a separate high-precision
      calculation of the $8s$   polarizability using the all-order
      matrix elements from Table~\ref{tab1}. We find  DHF result to be in
      good agreement with the high-precision value. 
         
    \begin{table}
\caption{\label{tab8} Dynamic polarizabilities 
$\alpha_{ns}(\omega)$ (in units of $a_0^3$) for Rb, 
$\omega$=0.0576645~a.u., i.e. $\lambda = 790$ nm. } 
 \begin{ruledtabular}
 \begin{tabular}{cccccc}
\multicolumn{1}{c}{$n$}& 
\multicolumn{1}{c}{$\alpha_{\mathrm{DHF}}$}& 
\multicolumn{1}{c}{$\alpha$}&
\multicolumn{1}{c}{}&
\multicolumn{1}{c}{$n$}& 
\multicolumn{1}{c}{$\alpha_{\mathrm{DHF}}$}\\ 
    \hline                              
    8   & -304  &-295\footnotemark[1]&  & 14   & -286  \\
    9   & -292  &                    &  & 15   & -285  \\
   10   & -289  &                    &  & 16   & -284  \\  
   11   & -288  &                    &  & 17   & -282\\
   12   & -287  &                    &  & 18   & -280\\
   13   & -287  &                    &  & 19   & -277\\
    \end{tabular}
\end{ruledtabular}
\footnotetext[1] {\noindent High-accuracy value obtained  using 
  experimental energies and  all-order matrix 
  elements for the dominant terms with $n^{\prime}=7,~8$.}
\end{table} 
            
   \begin{figure}  [ht]
\includegraphics[scale=0.45]{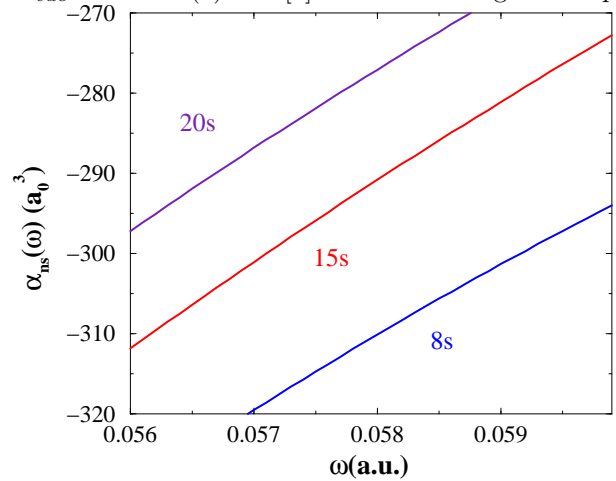}   
\caption{ \label{fig2} Dynamic polarizability $\alpha(\omega)$ for 
the $ns$ states  of Rb in $a^3_0$.  }             
\end{figure} 

   We plot the frequency-dependent polarizabilities of $8s$, $15s$, and
$20s$ 
  states  in Fig.~\ref{fig2}. As expected, the polarizabilities depend
   weakly on $\omega$ for the frequencies considered here. 
  The polarizabilities of the ground and Rydberg states can be matched 
   at the point between two resonances where the ground state polarizability 
   is small and negative. The exact matching point for the $15s$ state is 
   $\lambda_{vac}=790.14(2)$~nm \cite{us}. The matching of the 
   polarizabilities allows to minimize motional heating in the 
   quantum computation scheme with neutral atoms. 
                           \section{Conclusion}
We have conducted a systematic study of the $ns-n'p$ and $nd-n'p$
electric-dipole matrix elements
in rubidium using relativistic all-order method. An investigation of the
accuracy of these matrix elements was performed.
The resulting matrix elements were used to calculate lifetimes of the 
$ns$ and $np$ levels with $n=6,7,8$ and $nd$ levels with $n=4,5,6$.
The lifetime values were found to be in good agreement with experiment.
The dynamic polarizabilities of the $ns$ Rb states, which are of interest
for 
  the  optimization of quantum computation scheme with neutral atoms
 mediated by the conditional excitations to Rydberg states, were  
 also
 calculated. 
  \begin{acknowledgments}
   This work was partially supported by the Advanced
Research Development Activity, the National Security Agency, 
and NIST Advanced Technology Program. 
 \end{acknowledgments}


\begin{thebibliography}{21}
\bibitem[{wav()}]{wavelength_comment}
\bibinfo{note}{We use the value $\omega=0.04298$~a.u. that was used in
  Ref.~\cite{msd}, but we note that the value $\omega=0.04281$~a.u. actually
  corresponds to the wavelength in air of $\lambda=1.064 \mu$m of the Nd:YAG
  laser.}
\end{thebibliography}
\end{document}